\documentstyle[12pt]{article}

\newcommand{\EQ}{\begin{equation}}
\newcommand{\EN}{\end{equation}}
\newcommand{\bea}{\begin{eqnarray}}
\newcommand{\ena}{\end{eqnarray}}
\newcommand{\vs}[1]{\vspace{#1 mm}}

\newcommand{\uda}{\nearrow \kern-1em \searrow}

\def\ket#1{\vert #1>}
\def\vac{\ket 0}

\def\oot{{ 1 / 2} }

 \def\qh{{q^{1/2}}}
 \def\qmh{{q^{-1/2}}}
 \def\qq{{q^{1/4}}}
 \def\qmq{{q^{-1/4}}}
 \def\qhmh{{(\qh - \qmh)}}

 \def\cone{c_1}
 \def\ctwo{c_2}
 \def\cthree{c_3}
 \def\cfour{c_4}
 \def\bone{b_1}
 \def\btwo{b_2}
 \def\bthree{b_3}
 \def\bfour{b_4}
\begin{document}

\topmargin 0pt
\oddsidemargin 5mm

\renewcommand{\Im}{{\rm Im}\,}
\newcommand{\NP}[1]{Nucl.\ Phys.\ {\bf #1}}
\newcommand{\PL}[1]{Phys.\ Lett.\ {\bf #1}}
\newcommand{\CMP}[1]{Comm.\ Math.\ Phys.\ {\bf #1}}
\newcommand{\PR}[1]{Phys.\ Rev.\ {\bf #1}}
\newcommand{\PRL}[1]{Phys.\ Rev.\ Lett.\ {\bf #1}}
\newcommand{\PTP}[1]{Prog.\ Theor.\ Phys.\ {\bf #1}}
\newcommand{\PTPS}[1]{Prog.\ Theor.\ Phys.\ Suppl.\ {\bf #1}}
\newcommand{\MPL}[1]{Mod.\ Phys.\ Lett.\ {\bf #1}}
\newcommand{\IJMP}[1]{Int.\ Jour.\ Mod.\ Phys.\ {\bf #1}}

\begin{titlepage}
\setcounter{page}{0}
\begin{flushright}
OU-HET 200\\
\end{flushright}

\vs{15}
\begin{center}
{\Large CHIRAL FERMIONS\\
ON QUANTUM FOUR-SPHERES}
\vs{15}

{\large Kazutoshi Ohta
\footnote{e-mail address: kohta@phys.sci.osaka-u.ac.jp}
and Hisao Suzuki
\footnote{e-mail address: suzuki@phys.wani.osaka-u.ac.jp}}\\
\vs{8}
{\em Department of Physics, \\
Osaka University \\ Toyonaka, Osaka 560, Japan} \\
\end{center}
\vs{10}

\centerline{{\bf{Abstract}}}
We construct wave functions and Dirac operator of spin $1/2$ fermions
on quantum four-spheres. The construction can be achieved by the
q-deformed differential calculus which is manifestly
$SO(5)_q$ covariant. We evaluate the engenvalue of the Dirac
operator on  wave functions of the spinors and show  that we can define the
chiral fermions in such a way that  the massless Dirac
operator anti-commutes with $\gamma_5$.

\end{titlepage}
\newpage
\renewcommand{\thefootnote}{\arabic{footnote}}
\setcounter{footnote}{0}

The structure of quantum groups\cite{Dri}\cite{Jim} has been appeared
in many aspect of
the field theories. As geometrical counterparts of quantum groups, we
have
non-commutative geometry\cite{Wor}\cite{Man}.
The quantum manifolds have been investigated by physician as possible
deformations of the space-time\cite{Pod}\cite{WZ}. However, up to now
there is no priori
reason why we have to consider quantum manifolds. It seems of no
use to consider quantum manifolds instead of classical
manifolds. As an attempt to
find a reasonable solution,  the use of
quantum spheres as a regularization of field theories was
proposed~\cite{Suzuki}.
Actually, it is shown that the number of wave functions of the scalar
fields on quantum spheres can be finite
when $q$ is a root of unity by using periodic representations. The
parameter is determined by the
radius of the spheres and cutoff scale. At low energy, the observable
 reduces to the ones on classical spheres.  We can thus regard the
 quantum manifold as an effective prescription of the strong quantum
 fluctuation of space-time at microscopic level.  Of course,
the investigation of quantum field theories on quantum spheres was far
from complete. The
 calculation of the amplitudes or full second quantization are
 importaint problem which we should solve for the  understanding of the
field
 thoeries on quantum spheres. Before analyzing these questions, we
 should first construct the fields with internal spin on quantum
spheres.

Our aim of this paper is to construct spin $1/2$ fermions and dirac
operators on quantum four-spheres.  Since we cannot  change   variables
freely on quantum
spheres, it is necessary to use a formulation which is manifestly
covariant under the quantum groups.  For
classical four-spheres, there exist such a formulation. The manifest
covariant formulation of the spinors and gauge fields was formulated
long ago by
Adler~\cite{Adl}. The instanton-(ant-instanton) solution was shown to
be
 expressed elegantly in ref.\cite{JR}. We therefore  consider the
quantum
 version of the formulation.  In the case of quantum two-spheres, we
 have already found the dirac operators\cite{OS}. The advantage of this
formulation is that we can easily consider the higher dimensional
extension, which we are going to pursue.

The generators of quantum $SO(5)$\cite{Lus} satisfy the following
relations:
\bea
K_i X_j^\pm K_i^{-1} &=& q_i^{ \pm a_{ij} } X_j^\pm, \nonumber\\
\left[ {X_i}^{+}, {X_j}^{-} \right] &=& \delta_{ij}{ K_i - K_i^{-1}
\over q_i - q_i^{-1}},\nonumber\\
\sum_{\nu=0}^{1-a_{ij}}&(-1)^\nu& \left[ \matrix{1-a_{ij} \cr
  \nu\cr}\right]_{q_i} (X_i^\pm)^{1-a_ij-\nu}X_j^\pm(X_i^\pm)^\nu
=0\quad(i\neq j), \nonumber\\
q_1 &=& q, \qquad q_2 = q^{1/2}.
\ena
The co-product rules we will use are
\bea
\Delta(X_i^\pm) = X_i^\pm \otimes K_i^{-1/2} + K_i^\oot \otimes
X_i^{\pm}, \qquad
\Delta(K_i) = K_i \otimes K_i.
\ena

We define the coordinates of the four-sphere as a $5$ dimensional
representation

of $SO(5)_q$\cite{YRTF}\cite{Tak}, which are denoted by $e_m$
$(m=1,2,\ldots,5)$. The action of the

generators $X_i^\pm$ and $K_i$ are given by
\bea
\pi(X_1^+) &=& E_{12} - E_{45},\qquad \pi(X_1^-) = E_{21} - E_{54}, \\
\pi(K_1) &=& q^{E_{11} - E_{22} + E_{44} - E_{55}}, \\
\pi(X_2^+) &=& [2]^{1/2} ( E_{23} - E_{34} ), \qquad \pi(X_2^-) =
[2]^\oot (E_{32} - E_{43}), \\
\pi(K_2) &=& q^{E_{22} - E_{44}},
\ena
where $q$-integers $[n]$ is defined by
\bea
[n] = { q^{n/2} - q^{-n/2} \over \qh - \qmh }.
\ena
The commutation relations can be defined by demanding that

$15$ dimensional representation can not be constructed by the product
of
the coordinates:
\bea
e_i e_j &=& q e_j e_i, \quad i<j,\quad i \neq j' , \\
e_2e_4 &-& e_4 e_2 + \qhmh e_3^2 = 0 \\
e_1e_5&=& e_5 e_1 + \qhmh ( \qh e_4 e_2 + \qmh e_2 e_4 + e_3^2) = 0,
\ena
where $j' =6-j$.
The equation of the sphere is given by
\bea
r^2 = q^{-3/2} e_1 e_5 + \qmh e_2 e_4 + e_3^2 + \qh e_4 e_2 + q^{3/2}
e_5 e_1.
\ena
The states whose highest weight state is given by $e_1^l$ forms an

${1 \over 6} (l+1)(l+2)(2l+3)-$dimensional representation.

We introduce $4$ dimensional representation of $SO(5)_q$ given by

$c_m$ $m=1,\ldots, 4)$. The action of Generators $X_i^\pm$ and $K_i$
can
be found as

\bea
\pi(X_1^+) &=& E_{23}, \qquad \pi(X_1^-) = E_{32}, \\
\pi(K_1) &=& q^{E_{22} - E_{33}}, \\
\pi(X_2^+) &=& E_{12} - E_{34}, \qquad \pi(X_2^-) = E_{21} - E_{43},\\
\pi(K_2) &=& q^{(E_{11} - E_{22} + E_{33} - E_{44})/2},
\ena
which is the action of $SP(4)_{q^{1/2}}$ on its fundamental
representations.
We define the derivatives with respect to the coordinates by
\bea
\partial_i e_j = q^{-1} (R^{5,5})_{ij}^{kl} e_l \partial_k + q^{\rho_j}
\delta_{i,j' },
\ena
where $\rho_i = (3/2, 1/2,0,-1/2,-3/2)$, and $(R^{5,5})_{ij}^{kl}$ is
the R-matrix\cite{Jim2}\cite{KR}:
\bea
R^{5,5} &=& q\sum_{\scriptstyle i=1\atop\scriptstyle i \ne 3}^{5}
e_{ii}\otimes e_{ii} + e_{33}\otimes e_{33} + \sum_{\scriptstyle i,j =
  1\atop\scriptstyle i \ne j,j' }^5 e_{ii} \otimes e_{jj} \nonumber \\
& & +q^{-1} \sum_{\scriptstyle i=1\atop\scriptstyle i \ne i' }^{5} + (
q
- q^{-1}) \sum_{\scriptstyle i,j = 1\atop\scriptstyle i > j}^5 e_{ij}
\otimes e_{ji} \nonumber\\
& & -(q-q^{-1}) \sum_{\scriptstyle i,j = 1 \atop\scriptstyle i>j}^5
 q^{\rho_i - \rho_j} e_{ij}e_{i' j' },
\ena

As for the conjugate for the spinor, we impose the following condition:
\bea
b_j c_i \vac = \varepsilon_i q^{\rho_i} \delta_{i,j\prime } \vac,
\ena
where $\rho_i = ( 1, 1/2, -1/2, -1)$,$\varepsilon_i = (1,1,-1,-1)$ and
$j\prime =5-j$.
We prepare another conjugate $b'$, which satisfy the  relations
identical to
(18) for $c_i$ but not for $e_i$ , which we will specify later.

We can define the commutation relations between them, but we do not
have to
use these relations in this paper because we only consider the one particle
states for the spinors.

As  commutation relations between $e_i$ and $c_i$, we take
\bea
c_i e_j = (R^{4,5})_{ij}^{kl} e_l c_k,
\ena

where $R^{4,5}$ is the R-matrix between the $4$ and $5$ dimensional
representations. We can find the explicit form by the following
procedure.
At first, we can find the $R$-matrix between $4$ dimensional
representations
from ref.\cite{YRTF} as
\bea
R^{4,4} &=& q^{1/2}\sum_{ i=1}^{4}
e_{ii}\otimes e_{ii} + \sum_{\scriptstyle i,j =
  1\atop\scriptstyle i \ne j,j\prime }^4 e_{ii} \otimes e_{jj}
\nonumber \\
& & +q^{-1/2} \sum_{i=1}^{4} + ( q^{1/2}
- q^{-1/2}) \sum_{\scriptstyle i,j = 1\atop\scriptstyle i > j}^4 e_{ij}
\otimes e_{ji} \nonumber\\
 & & -(q^{1/2}-q^{-1/2}) \sum_{\scriptstyle i,j = 1 \atop\scriptstyle
i>j}^4
 q^{\rho_i - \rho_j} \varepsilon_i \varepsilon_j e_{ij}e_{i\prime
j\prime },
\ena
We next construct $5$ dimensional representations by tensor product of
$4$ dimensional representation as
\bea
\ket{5,1} &=& {1 \over [2]^{1/2}}( \qmq \ket{4,1} \otimes \ket {4,2} -
  \qq \ket{4,2}\otimes \ket{4,1}),\nonumber\\
\ket{5,2} &=& {1 \over [2]^{1/2}}( \qmq \ket{4,1} \otimes \ket {4,3} -
  \qq \ket{4,3}\otimes \ket{4,1}),\nonumber\\
\ket{5,3} &=& {1 \over [2]}( \qmh \ket{4,2} \otimes \ket {4,3} -
  \qh \ket{4,3}\otimes \ket{4,2} - \ket{4,1}\ket{4,4} +
\ket{4,4}\ket{4,1}
),\nonumber\\
\ket{5,4} &=& {1 \over [2]^{1/2}}( \qmq \ket{4,2} \otimes \ket {4,4} -
  \qq \ket{4,4}\otimes \ket{4,2}),\nonumber\\
\ket{5,5} &=& {1 \over [2]^{1/2}}( \qmq \ket{4,3} \otimes \ket {4,4} -
  \qq \ket{4,4}\otimes \ket{4,3}),
\ena
Then we can find the $R$-matrix between the $4$-dimensional
representations and $5$-dimensional representations  by successively
applying the formula $(20)$.

As for the commutation relations for $b_i$, we use
\bea
b_i e_j = ((R^{4,5})^{-1})_{ij}^{kl} e_l b_k,
\ena
and for $b'_i$, we choose
\bea
b_i '  e_j = (R^{4,5})_{ij}^{kl} e_l b_k ' .
\ena
The $q$-analog of $\gamma$ matrices can be defined by the tensor
product
of $c_i$ and $b_j$ as
\bea
\gamma_1 &=&  \qmq c_1 \otimes b_2 -
  \qq c_2\otimes b_1,\nonumber\\
\gamma_2 &=&  \qmq c_1 \otimes b_3 -
  \qq c_3 \otimes b_1,\nonumber\\
\gamma_3 &=& {1 \over [2]^{1/2}} (\qmh c_2 \otimes b_3 -
  \qh c_3 \otimes b_2 - c_1 b_4 + c_4 b_1
),\nonumber\\
\gamma_4 &=&  \qmq c_2 \otimes b_4 -
  \qq c_4 \otimes b_2 ,\nonumber\\
\gamma_5 &=&  \qmq c_3 \otimes b_4 -
  \qq c_4 \otimes b_3,
\ena
We define the operator of orbital angular momentum by constructing
$10$-dimensional representation by tensor product of $e_i$ and
$\partial_j$;
\bea
J_{ij} &=& \qmh e_i \Lambda \partial_j - \qh e_j \Lambda \partial_i ,
\quad i<j,\quad i \neq j\prime , \nonumber\\
J_{24} &=& e_2 \Lambda \partial_4 - e_4 \Lambda \partial_2 + \qhmh e_3 \Lambda
\partial_3, \nonumber\\
J_{15} &=& \Bigl({[2][3] \over [6]}\Bigr)^{1/2}\bigr[ e_1 \Lambda
\partial_5
- e_5 \Lambda \partial_1
+ (q-1)e_4 \Lambda \partial_2 \nonumber\\
& & +(1-q^{-1})e_2
\Lambda \partial_4 + \qhmh e_3 \Lambda \partial_3\bigl],
\ena
where $\Lambda$ is the dilatation operator\cite{Ogi}\cite{OZ}\cite{Fio}
satisfying
\bea
\Lambda e_i = q e_i \Lambda.
\ena
The introduction of this operator is required to ensure relations;
\bea
[J_{ij}, r^2] = 0.
\ena

We define the $q$-analog of spin matrix $S_{ij}$ as
\bea
S_{12} &=& [2] \cone \bone,\qquad S_{13} = [2]^{1/2}( \qmq \ctwo \bone
+
\qq \cone \btwo), \nonumber\\
S_{14} &=& - [2] \ctwo \btwo, \qquad S_{23} = [2]^{1/2}(\qmq \cthree
\bone + \qq \cone \bthree),\nonumber\\
S_{25}&=& [2] \cthree\bthree, \qquad S_{34} = [2]^{1/2}(\qmq
\cfour\btwo
+ \qq \ctwo\bfour),\qquad S_{45}= - [2] \cfour \bfour,\nonumber\\
S_{24}&=& \qmh \cfour\bone + \qh\cone\bfour -
\cthree\btwo - \ctwo \bthree,\nonumber\\
S_{15}&=& \Bigl({ [2][3] \over [6]}\Bigr)^{1/2}(\qmh \cfour\bone + \qh
\cone\bfour + q^{-1}\cthree\btwo + q\ctwo\bthree),
\ena
and singlet operator $O$ as
\bea
O = q^{-1}\cone\bfour + \qmh \ctwo\bthree - \qh \cthree\btwo - q
\cfour\bone.
\ena
We will also use $S_{ij}'$ and $O'$ which are defined just by replacing

$b$ in $(28)$ and $(29)$ by $b'$.

These are all the relations we use to construct Dirac operators.

{}From the tensor product of ${1 \over6}(l+1)(l+2)(2l+3)$dimensional
representation
of the coordinates and $4$ dimensional spinor, we have $g_l$ and
$g_{l-1}$
dimensional representations where $g_l = {2 \over
3}(l+1)(l+2)(l+3)$\cite{CW}.
The highest weight states are given by
\bea
\ket{l,g_l,h} &=& a_l e_1^l c_1 \vac, \nonumber\\
\ket{l,g_{l-1},h} &=& a_l \Bigl( {[2l] \over [2l+3]}\Bigr)^{1/2}
( [2]^{-1/2} q e_1^{l-1}e_3c_1 - \qq e_1^{l-1}e_2 c_2 \nonumber\\
& & +q^{-3/4}e_1^l c_3) \vac,
\ena
where $a_l$ is a constant which can be specified by haar measure.
In the previous paper, we have shown that the structure of the
currents
for quantum 2-sphere\cite{OS}
can be described in terms of C-G coefficients\cite{KR}.  We can follow
the
similar argument for this four spheres. However,
in the case of $SO(5)_q$, it seems not convenient to use C-G
coefficients
because
we do not have any economical expression of the C-G coefficients.
Therefore,
we directly consider the singlet operators (Casimir operators), so
that the effect on the states can
be analyzed just by applying them to the highest weight states $(30)$.

The eigenvalues of a singlet operator $ J \cdot S$ on the highest
weight representations are given, after tedious
but straightforward calculation, by
\bea
J\cdot S \ket{l,g_l,h} &=& { [3][2l] \over [6]}q^{l+3}
\ket{l,g_l,h},\nonumber\\
J\cdot S \ket{l,g_{l-1},h} &=& { [3][2l+6] \over [6]}q^{-l}
\ket{l,g_{l-1},h},
\ena
We also find

\bea
O \ket{l,g_l,h} &=& q^{l}
\ket{l,g_l,h},\nonumber\\
O\ket{l,g_{l-1},h} &=& q^{-l-3}
\ket{l,g_{l-1},h},
\ena
We also define the $\gamma_5$ by
\bea
\gamma_5 = {c \over r} e \cdot {\gamma}',
\ena
where the coefficient $c$ can be fixed by requiring $\gamma_5^2 = - 1$.

This operator changes the representation of the coordinate but not the
total spin. We therefore have
\bea
\gamma_5 \ket{l,g_l,h} &=& \alpha_l
\ket{l+1,g_l,h},\nonumber\\
\gamma_5\ket{l,g_{l-1},h} &=& \beta_l
\ket{l-1,g_{l-1},h},
\ena
where $\alpha_l$ and $\beta_l$ are some coefficients which we do not
need to evaluate for our present purpose.

We can find that the following form of the operator anti-commute with
$\gamma_5$:
\bea
\gamma \nabla \equiv {1 \over r} O^{-1}\Bigl( q^{-1} J \cdot S  +
{[3][4] \over [6]}\Bigl),
\ena
where $r$ is the radius of the sphere appeared in $(11)$.
The eigenvalue of this operator for the states is given by
\bea
\gamma \nabla \ket{l,g_l,h} &=& {[3][2l+4] \over [6]r}
\ket{l,g_l,h},\nonumber\\
\gamma \nabla \ket{l,g_{l-1},h} &=& -{[3][2l+2] \over [6]r}
\ket{l,g_{l-1},h}.
\ena

We have shown that Dirac operators can be constructed by a manifest
$SO(5)_q$ invariant method, quite analogous to the classical
case.\cite{Adl}
For classical spheres, it was shown that the gauge fields can be
described quite elegantly and instanton-(anti-instanton) solutions
have  a simple
form\cite{JR}. We can therefore expect that such
solutions can be
obtained also for quantum spheres.
These considerations might be interesting when we consider the topology
of the quantum manifold.

\end{document}